\documentstyle[11pt,epsf,mypp]{article}

\newcommand{\comment}[1]{}

\newcommand{\myshowfig}[1]{#1}

\def\be{\begin{equation}}
\def\bel#1{\begin{equation}\label{eq:#1}}
\def\ee{\end{equation}}
\def\bea{\begin{eqnarray}}
\def\beal#1{\begin{eqnarray}\label{eq:#1}}
\def\eea{\end{eqnarray}}
\def\eqref#1{\ref{eq:#1}}

\def\eqn#1{equation~(\ref{eq:#1})}

\def\fig#1{Figure~\ref{fig:#1}}

\def\lci{LC1}
\def\lcii{LC2}

\def\etal{~\mbox{et al.}}

\def\kms{km/s}

\def\Mstar{M_{\ast}}

\def\gsim{\lower 2pt \hbox{$\, \buildrel {\scriptstyle >}\over {\scriptstyle
\sim}\,$}}
\def\lsim{\lower 2pt \hbox{$\, \buildrel {\scriptstyle <}\over {\scriptstyle
\sim}\,$}}

\begin{document}

\title{Density-Dependent Luminosity Functions for Galaxies in the 
       Las Campanas Redshift Survey}

\author{
Benjamin C. Bromley \& William H. Press, 
}
\affil{
Physics Department, Harvard University, \& 
Harvard-Smithsonian Center for Astrophysics
}
\authoraddr{
60 Garden Street, Cambridge, MA 02138
}
\author{
Huan Lin
}
\affil{
Department of Astronomy, University of Toronto
}
\authoraddr{
60 St. George Street, Toronto, ONT M5S 3H8, Canada
}
\centerline{and}
\author{
Robert P. Kirshner
}
\affil{
Harvard-Smithsonian Center for Astrophysics
}
\authoraddr{
60 Garden Street, Cambridge, MA 02138
}

\begin{abstract}

Galaxies in the Las Campanas Redshift Survey are classified according
to their spectra, and the resulting spectral types are analyzed to
determine if local environment affects their properties.  We find that
the luminosity function of early-type objects varies as a function of
local density. Our results suggest that early-type galaxies
(presumably ellipticals and S0's) are, on average, fainter when they
are located in high-density regions of the Universe. The same effect
may operate for some, but not all, late-type galaxies. We discuss the
implications of this result for theories of galaxy formation and
evolution.

\end{abstract}


\section{Introduction}\label{sect:intro}

As a snap-shot of a dynamic, evolving system, the present-day
distribution of galaxies offers strong constraints for theories of
structure formation. One of the best established of these, the
density-morphology relation (Dressler~\cite{Dre80}), quantifies the
extent to which the mix of galaxy types is dependent on the local
environment.  It is likely that the principal process responsible
for this relation is the merging of galaxy pairs: Elliptical
galaxies lie preferentially in high-density regions because, there, they
had access to an abundant supply of objects with which to merge.

Any galaxy-formation scenario must also account for the observations
of galaxies at earlier epochs (e.g., Ellis \cite{Ell98}).  For
example, a population of late-type galaxies observed at $z \approx 1$
has no apparent counterpart among galaxies seen in the present epoch
(Broadhurst, Ellis \& Shanks~\etal\ \cite{BroEtal88}). Conversely,
some early-type (S0) objects are relatively abundant at present, but
similar objects, or even obvious progenitors, are less prevalent at a
redshift near unity (Dressler~\etal\ \cite{DreEtal97}).  It may not be
simple to connect galaxies of the distant past and those of the
present epoch with a consistent model of cosmic structure formation.

The purpose of this note is to report an effect which adds to the
growing list of requirements for a successful theory of galaxy
formation. Specifically, in a study of a population of galaxies in the
present epoch, we find that the luminosity function of some galaxy
types depends on local density: Early-type galaxies tend to be fainter
when they are located in dense environments.  This result is based on
an analysis of galaxies in the Las Campanas Redshift Survey
(Shectman~\etal\ \cite{SheEtal96}), the parameters of which are
briefly outlined here in \S 2.  We use a spectral classification
scheme and a simple partition of galaxies into low- and high-density
subsamples to evaluate the environmental dependence of type-specific
luminosity functions; details and specific results are given in \S 3.
We finish this note with possible interpretations of our results (\S
4).

\section{The Redshift Data}\label{sect:data}

The Las Campanas Redshift Survey consists of approximately 26,000
galaxies selected in the $R$-band, as described by Shectman~et~al.\
(\cite{SheEtal96}). The survey volume consists of six narrow strips in
the plane of the sky, three in the northern Galactic hemisphere and
three in the south, which an effective depth of roughly 30,000~\kms.
The selection of galaxies is somewhat detailed, with apparent
magnitude limits, central surface brightness cuts, and sampling rates
that vary across the sky.  All of the selection parameters have been
quantified and are incorporated into the analysis performed here
following the prescription given by Lin~et~al.\ (\cite{LinEtal96}).
As in the previous work, when estimating luminosity functions we limit
ourselves to regions of the sky which were sampled at with a 112-fiber
spectrometer and which have the largest sampling rates and least
stringent central surface brightness cuts.

Here we identify galaxy types using the spectral classification scheme
described by Bromley \etal\ (\cite{BroEtal98}). Types come from a
principal component analysis of rest-frame galaxy spectra and are
distinguished largely by the degree of line emission superimposed on a
spectrum typical of a red stellar population.  We consider the same six
types defined in the earlier paper, labeled $C = {1,2,...,6}$;
$C = 1$ objects have spectra with no emission-line features, while
$C = 6$ galaxies show very strong emission.  The correlation between
the spectral types and morphology is expected to strong (e.g.,
Kennicutt \cite{Ken92}) and the nomenclature of ``early'' (connoting ellipticals and S0's) and ``late'' (connoting spirals)
are applied to low and high $C$ values, respectively.

\section{Luminosity Functions by Type and Density}

Our central result is a variation in the luminosity distribution of
some galaxy types with environment. The analysis is essentially
statistical; we consider only the distribution of luminosity among
galaxies of a given type.  Here luminosity functions are estimated in
two ways, following Lin~et~al.\ (\cite{LinEtal96}).  The first is the
non-parametric estimator proposed by Efstathiou, Ellis \&
Peterson~(\cite{EfsEllPet88}) to obtain the unconstrained form of the
luminosity functions.  The second is a fit to a Schechter function,
\bel{schechter}
\Phi(M) = A 10^{+0.4 (\alpha+1) (\Mstar-M)} \exp(-10^{0.4 (\Mstar - M)}) 
\ ,
\ee
where $\alpha$ and $\Mstar$ correspond to the slope of the faint-end
of the galaxy distribution and to a typical absolute magnitude,
respectively, and $A$ is a normalization constant.  To obtain both
types of estimators, we use a maximum likelihood method (Sandage,
Tammann \& Yahil \cite{SanTamYah79}).  In the case of the Schechter
parameterization, we limit the range in absolute R-band magnitude to
$-23 < M_r < -17.5$, as in Lin~et~al.\ (\cite{LinEtal96}).

We must also devise a way for determining the local density of a
galaxy so that the survey can be partitioned into high and low density
regions. For this purpose a friend-of-friends algorithm (Huchra \&
Geller \cite{HucGel82}) is sufficient, as the objective here is to
simply and unambiguously determine high density regions above some
threshold.  An advantage of this method is that prominent
redshift-space distortions in virialized clusters are handled quite
well. A disadvantage is that the method is susceptible to
misclassification of galaxies in small groups.  Here we choose link
parameters of 75~km/s in the plane of the sky and 500~km/s along the
line of sight. The resulting grouped galaxies, tagged as
``high-density'', with a 4-object minimum group size, constitute about
one-third of the total survey.  A crude estimate of the density
threshold is approximately 1,000 times the mean.  Galaxies observed with
the 50-fiber spectrometer were included in the group-finding step so
that dense structures can span the gaps in the survey which may exist
between the 112-fiber fields.

\begin{figure}[h]
\myshowfig{\epsfxsize=6.5in\epsfbox{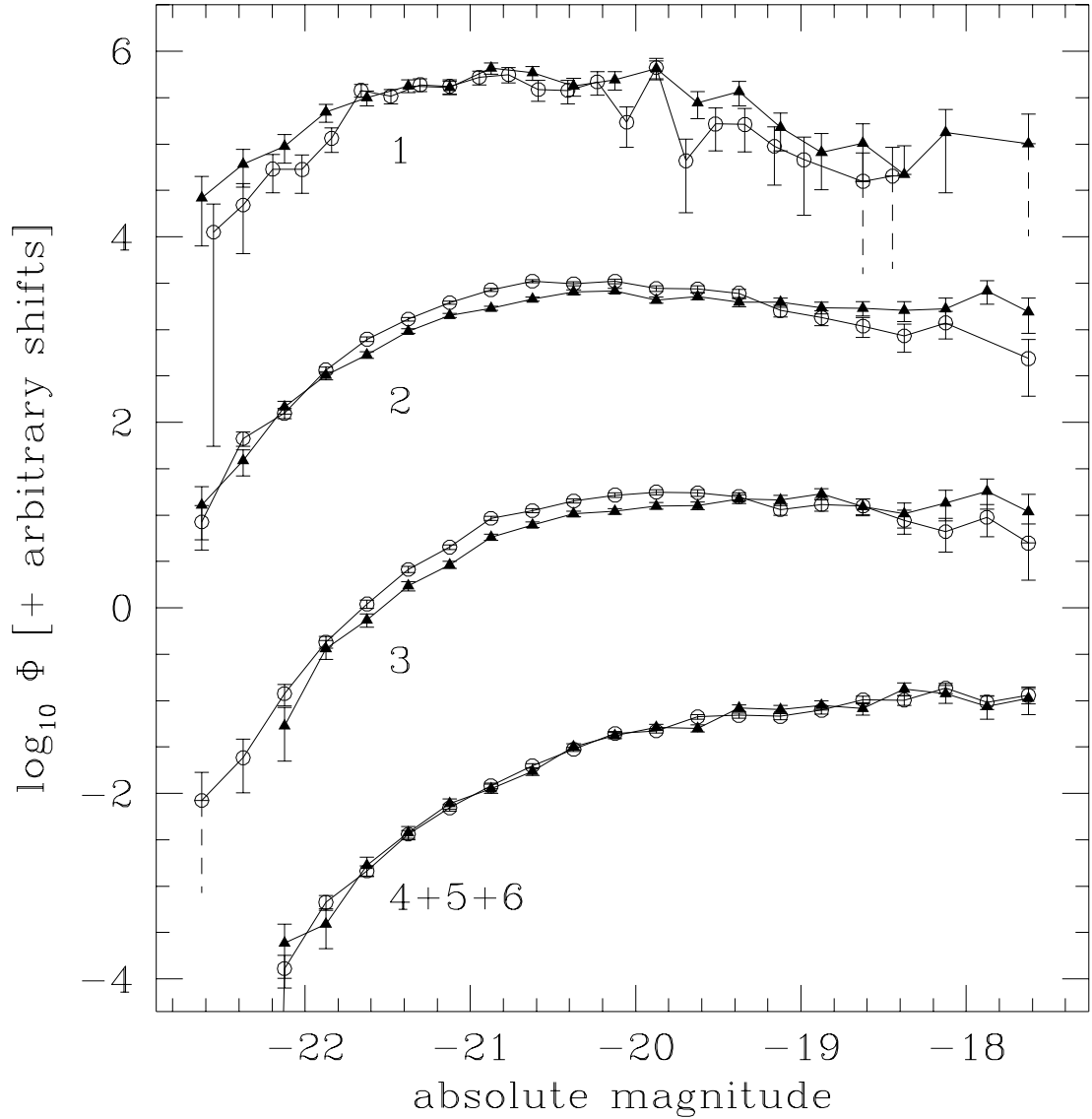}}
\caption{The luminosity function by type and density.  Nonparametric
fits are shown for objects in high-density regions (dark triangles)
and low-density regions (open circles). The labels indicate type
index; the pair of curves at the bottom were produced by merging
samples of the three late-type objects.  The excess numbers of faint,
early-type galaxies in the high-density cases are evident from the
relative flatness of the luminosity functions above a magnitude of
--20. The effect is not significant for the combined late-type galaxies.}
\label{fig:hilolfs}
\end{figure}

\begin{figure}[h]
\myshowfig{
\epsfxsize=6.5in\epsfbox{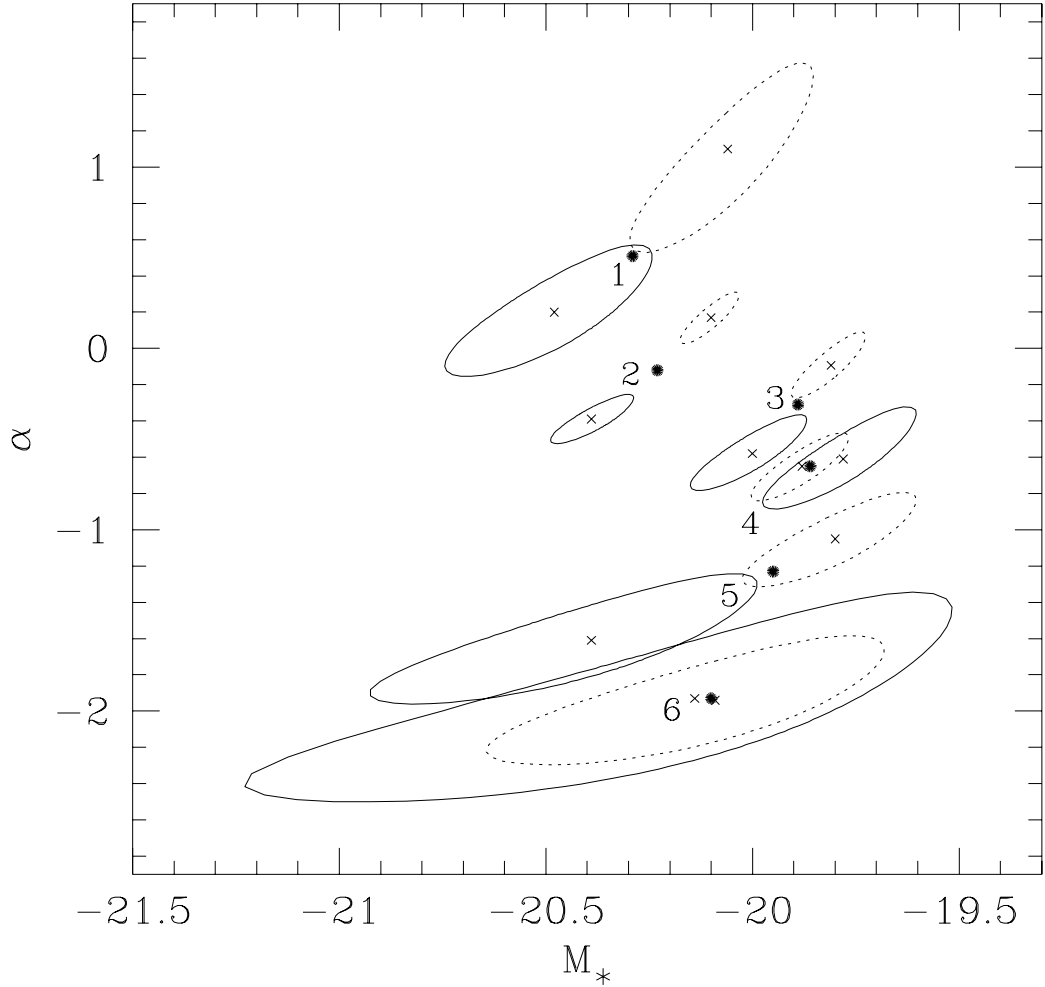}
}
\caption{Error ellipsoids for the best-fit Schechter parameters for
objects in high and low density regions.  The ellipsoids show 95\%
confidence intervals for objects found in high-density regions (solid
contours) and low-density regions (dashed contours). The labels
indicate type index, the $\times$ symbols mark best-fit values for the
subsamples, and the solid circles indicate best-fit
density-independent values.  There is a significant shift in both
$M_\ast$ and $\alpha$ for some objects, particular those of type 2, as
a result of local density. The downward translation of $\alpha$ values
indicates a relative excess in the number of faint galaxies in the
high-density cases.}
\label{fig:hilolfam}
\end{figure}

\fig{hilolfs} illustrates the dependence of the type-dependent luminosity
functions on density while \fig{hilolfam} gives the best-fit Schechter
parameters. The early-type galaxies ($C =  1,2,3$), accounting for over
70\% of the objects in the survey, show significant variation: the
faint-end slope steepens with density, with $\alpha$ shifting by about
--0.5.  From the best-fit values of the Schechter parameters, we
calculate that the mean luminosity of early-type galaxies in high-density
regions is approximately 50--80\% less than in the low-density
regions.  The late-type objects show little or no significant trend.

We must be wary that this result may reflect some property of the
survey other than true environmental dependence of type-specific
luminosity functions.  A flag for this cautionary statement is the
discrepancy in the {\em type-independent} luminosity functions
obtained in the northern and southern subsamples.  Lin~et~al.\
(\cite{LinEtal96}) found that the faint-end slopes from the Schechter
parameterization in these two regions differ by 0.1 ($\alpha_N =
-0.75\pm 0.04$ and $\alpha_S = -0.65\pm 0.04$). Lin~et~al.\ argue that
the mismatch is probably not entirely at the faint end, but represents
some broader differences in the luminosity function over the whole
range of magnitudes under consideration. 

Unfortunately, a north-south discrepancy persists in the type-specific
luminosity functions, although to a somewhat lessor degree. It is
statistically significant (above the 2-$\sigma$ level) only for types
1 and 3. Nonetheless, our result concerning the change of faint-end
slope with environment can be seen in the northern and southern
subsamples separately, and at about the same strength in terms of
changes in the faint-end slope.

We note that the type-independent luminosity functions of the high-
and low-density subsamples also exhibit differences in their faint-end
slopes. Compared to the full catalog, with $\alpha = -0.70\pm 0.03$,
we find values of $\alpha_h = -0.80\pm 0.05$ and $\alpha_l = -0.65\pm
0.05$ for high- and low-density objects respectively. The fact that 
the effect is weaker in the type-independent analysis is partially 
a reflection of the density-morphology relation---earlier types
are brighter on average than late types and are relatively more
numerous in high-density regions.

Other possible concerns include aperture effects from the finite size
of the spectrometer fibers ($3.5^{\prime\prime}$ in diameter).  These
could alter the type assignment of galaxies in a redshift-dependent
way if the spectrum from the central region of a galaxy differs
significantly from that of the galaxy as a whole. Presumably this
would affect late-type objects the most, since they may have localized
regions of star formation. One way to test for such a ``bias'' is to
partition the survey into high- and low-redshift subsamples and
estimate luminosity functions. Taking a threshold redshift of $cz =
30,000$~\kms, we find that there is a trend for faint-end slopes to be
steeper (more negative) in the high-$z$ subsample, but that the
difference is statistically significant for only one type ($C =
4$). Furthermore, within these subsamples, the environmental
dependence of $\alpha$ is observed, although it is stronger in the
low-$z$ galaxies.

We now consider the implications of these results for the overall
luminosity distribution of galaxies in the LCRS.  As recommended by
Binggeli, Sandage \& Tammann (\cite{BinSanTam88}), we may write the
general luminosity function of the full catalog as
\bel{genlf}
\Phi_{g}(M) = \sum_{c = 1}^{6} f_c \Phi_c(M) \ ,
\ee
where $\Phi_c$ is the luminosity function for type $c$, and $f_c$ is
the relative abundance of galaxies of this type in a given region of
the universe. This will be spatially varying as a result of the
density-morphology relation. The steepening of the luminosity function
with increasing density occurs simultaneously with the
density-morphology effect (see Table 1 of Bromley~\etal\
\cite{BroEtal98}). We can take both effects into account, writing a
more general form than in \eqn{genlf}, call it the grand luminosity
function,
\bel{grandlf}
\Phi_{G}(M) = \sum_{c,d} f_{cd} \Phi_{cd}(M) \ ,
\ee
where the fraction of galaxies $f$ and the type-dependent luminosity
functions both acquire a density index (here $d$ is a 1-bit number).
An estimate of this function from the LCRS data shows that it is not
significantly different from the general luminosity function defined
in \eqn{genlf}.  Thus for the purposes of understanding the global
properties of a catalog, the general luminosity function is a
reasonable approximation, as suggested by
Binggeli\etal~(\cite{BinSanTam88}).

\section{Discussion}

The density dependence of the faint-end slope of the type-specific
luminosity functions, call it the $\alpha$-density relation, suggests
that on average galaxies of a given type are fainter in dense regions
than in the field.  This behavior is qualitatively consistent with the
observation that the dwarf-to-giant ratio of early-type galaxies is
higher in groups than in the field (Ferguson \&
Sandage~\cite{FerSan91}).  However, from a theoretical standpoint,
this is perhaps an unexpected result.  Press-Schechter theory predicts
a relative decrease in low-mass objects in overdense regions, as a
result of a steeper decline of power with scale. One might assume that
this prediction applies to galaxies of a single type and therein lies
the unexpectedness of the $\alpha$-density relation.

It may be that the Press-Schechter prediction is more suggestive of
the change in the mix of galaxy types with density as opposed to the
dependence of intrinsic properties of any one galaxy type.  Thus, to
understand any sort of $\alpha$-density relation, one may have to
identify the processes which determine a galaxy's type.  Whatever
these processes, the LCRS data suggest that they allow galaxies of a
specific type to end up with less mass in dense regions than in the
field.  Toy models which show this behavior may be easy to construct,
but it may be more challenging to identify such an effect in more
developed theories of structure formation or numerical simulations.

Without subscribing to specific toy models, we note a few processes
which might affect the dependence of type-specific luminosity
functions. The most promising is tidal stripping of an evolved galaxy
or its progenitors in high-density regions. This would cause faint
galaxies to become fainter, and bright galaxies to become brighter,
assuming that the stripped material is transferred from one galaxy to
another.  There is a hint that this might be occurring: $\Mstar$
shifts slightly toward the bright end for galaxies in high-density
regions. Alternatively, the gaseous component of the stripped material
may be lost to the intergalactic or intracluster medium, and hence
would not strongly affect the bright population of galaxies.

Another possible process that could effect the $\alpha$-density
relation is quenching of star formation in high-density regions. If
one were to move a late-type galaxy from the field to a compact group
or cluster, it is possible that a hot intergalactic (or intracluster)
medium would produce sufficient $X$-radiation to reduce the star
formation rate in the galaxy. This would cause the galaxy to be
``misclassified'' as an earlier type in our spectral scheme. Such a
type migration would, to a crude approximation, steepen the faint-end
slope of an early-type population simply by mixing early- and
late-type luminosity functions.

One must be careful in associating the environmental dependence
of the luminosity function with changes in the brightness of
individual galaxies as predicted, for example, in the tidal stripping
scenario. In particular, making individual galaxies fainter does not
necessarily cause the slope of the luminosity function to rise toward the
faint end. If the faint-end slope is steep to begin with, and if
tidal stripping effects become more important with decreasing luminosity,
then the faint-end slope could actually decrease, at least when
measured over a finite range of luminosity. This might be a partial
explanation for the fact that two of the late-type luminosity functions
do not show a significant change in $\alpha$.

While the effect we have identified in the LCRS data may be a useful
constraint for theories of galaxy formation in its own right, we are
missing a possibly crucial body of information, namely a list of
galaxy morphologies.  The evidence is that morphology and spectral
type are strongly correlated, but not perfectly so (Kennicutt
\cite{Ken92}; Connolly \etal\ \cite{ConEtal95}; Zaritsky, Zabludoff \&
Willick \cite{ZarZabWil95}). Differences in this correlation from
high- to low-density regions might help to distinguish between the
various scenarios that could produce the environmental effect reported
here.

\acknowledgements

We thank Chuck Keeton, Chris Kochanek, and Ue-Li Pen for helpful
discussions.  Partial support for this work was provided by NSF Grants
PHY 95-07695 and AST 96-17058.  BCB is grateful to NASA Offices of
Space Sciences, Aeronautics, and Mission to Planet Earth for providing
computing resources.

\newpage

\end{document}